\documentclass[a4paper, 11pt,notoc]{article}
\pdfoutput=1
\usepackage{jcappub}
\usepackage{graphicx}
\usepackage{booktabs}
\usepackage{verbatim}
\usepackage{caption}
\usepackage{xspace}
\usepackage{hyperref}

\usepackage{subcaption}
\newcommand{\mDM}{\ensuremath{m_{\rm{DM}}}\xspace}
\newcommand{\mmed}{\ensuremath{M_{\rm{med}}}\xspace}

\newcommand{\gDM}{\ensuremath{g_{\rm{DM}}}\xspace}
\newcommand{\gq}{\ensuremath{g_q}\xspace}

\newcommand{\gdm}{\gDM}

\definecolor{cerulean}{RGB}{44,150,207}

\newcommand{\dmsimp}{\textsc{DMsimp}\xspace}
\newcommand{\maddm}{\textsc{MadDM}\xspace}

\RequirePackage[normalem]{ulem} %DIF PREAMBLE
\RequirePackage{color}\definecolor{RED}{rgb}{1,0,0}\definecolor{BLUE}{rgb}{0,0,1} %DIF PREAMBLE
\begin{document}
\title{\begin{boldmath} \huge Recommendations of the LHC Dark Matter Working Group: Comparing LHC searches for heavy mediators of dark matter production in visible and invisible decay channels\vspace{7mm} \end{boldmath}}

%%%%%%

\affiliation[*]{Editor}

\author[1,*]{Andreas~Albert,}
\affiliation[1]{III. Physikalisches Institut A, RWTH Aachen University, Aachen, Germany}

\author[2]{Mihailo Backovi\'c,}
\affiliation[2]{Center for Cosmology, Particle Physics and Phenomenology - CP3, Universite Catholique de Louvain, Louvain-la-neuve, Belgium}

\author[3,*]{Antonio~Boveia,}
\affiliation[3]{Ohio State University, 191 W. Woodruff Avenue
Columbus, OH 43210}
\emailAdd{antonio.boveia@cern.ch}

\author[4,*]{Oliver~Buchmueller,}
\affiliation[4]{High Energy Physics Group, Blackett Laboratory, Imperial College, Prince Consort Road, London, SW7 2AZ, United Kingdom}
\emailAdd{oliver.buchmueller@cern.ch}

\author[5,*]{Giorgio Busoni,} 
\affiliation[5]{ARC Centre of Excellence for Particle Physics at the Terascale, School of Physics, University of Melbourne, 3010, Australia}

\author[6,7]{Albert De~Roeck,}
\affiliation[6]{Antwerp University, BÐ2610 Wilrijk, Belgium. }
\affiliation[7]{CERN, EP Department, CH-1211 Geneva 23, Switzerland}

\author[8,*]{Caterina~Doglioni,}
\affiliation[8]{Fysiska institutionen, Lunds universitet, Lund, Sweden}
\emailAdd{caterina.doglioni@cern.ch}

\author[9,*]{Tristan~DuPree,}
\affiliation[9]{Nikhef, Science Park 105, NL-1098 XG Amsterdam, The Netherlands}

\author[10,*]{Malcolm Fairbairn,}
\affiliation[10]{Physics, King's College London, Strand, London, WC2R 2LS, UK}

\author[11]{Marie-H\'{e}l\`{e}ne Genest,}
\affiliation[11]{Laboratoire de Physique Subatomique et de Cosmologie, Universit\'{e} Grenoble-Alpes, CNRS/IN2P3, 53 rue des Martyrs, 38026 Grenoble Cedex, France}

\author[12]{Stefania Gori,}
\affiliation[12]{Department of Physics, University of Cincinnati, Cincinnati, Ohio 45221, USA}

\author[13]{Giuliano~Gustavino,}
\affiliation[13]{Universita' di Roma Sapienza, Piazza Aldo Moro, 2, 00185 Roma, Italy e INFN}

\author[14,*]{Kristian~Hahn,}
\affiliation[14]{Department of Physics and Astronomy, Northwestern University, Evanston, Illinois 60208, USA}
\emailAdd{kristian.hahn@cern.ch}

\author[15,16,*]{Ulrich~Haisch,}
\affiliation[15]{Rudolf Peierls Centre for Theoretical Physics, University of Oxford, Oxford, OX1 3PN, United Kingdom}
\affiliation[16]{CERN, TH Department, CH-1211 Geneva 23, Switzerland}
\emailAdd{ulrich.haisch@physics.ox.ac.uk}

\author[7]{Philip~C.~Harris,} 

\author[17]{Dan~Hayden,}
\affiliation[17]{Michigan State University, 220 Trowbridge Rd, East Lansing, MI 48824,
USA}

\author[18]{Valerio~Ippolito,} 
\affiliation[18]{Laboratory for Particle Physics and Cosmology, Harvard University, USA}

\author[8]{Isabelle~John,}

\author[19,*]{Felix~Kahlhoefer,}
\affiliation[19]{DESY, Notkestra\ss e 85, D-22607 Hamburg, Germany}

\author[20]{Suchita~Kulkarni,}
\affiliation[20]{Institut f\"ur Hochenergiephysik, \"Osterreichische Akademie der Wissenschaften, Nikolsdorfer Gasse 18, 1050 Wien, Austria}

\author[21]{Greg Landsberg,}
\affiliation[21]{Brown University, Dept. of Physics, 182 Hope St, Providence, RI 02912, USA.}

\author[22]{Steven Lowette,}
\affiliation[22]{Physics Department, Vrije Universieit Brussel, Brussels, Belgium}
\emailAdd{steven.lowette@cern.ch}

\author[11]{Kentarou~Mawatari,}

\author[23]{Antonio Riotto,}
\affiliation[23]{D\'{e}partement de Physique Th\'{e}orique (DPT) Geneva Cosmology and Astroparticle Group and Centre for Astroparticle Physics  (CAP), 24 quai Ernest Ansermet
CH-1211 G\'{e}n\`{e}ve}

\author[24]{William~Shepherd,}
\affiliation[24]{Institut f\"{u}r Physik, Johannes-Gutenberg-Universit\"{a}t Mainz, Staudingerweg 7, D-55128 Mainz, Germany}

\author[25,*]{Tim~M.P.~Tait,}
\affiliation[25]{Department of Physics and Astronomy, University of California, Irvine, California 92697, USA}
\emailAdd{ttait@uci.edu}

\author[3]{Emma Tolley,}

\author[10,*]{\\ Patrick Tunney,}

\author[26,*]{Bryan~Zaldivar,}
\affiliation[26]{Laboratoire de Physique Th\'{e}orique d'Annecy-le-Vieux (LAPTH) 9 Chemin de Bellevue, B.P. 110, F-74941 Annecy-le-Vieux, CEDEX, France}

\author[24]{Markus~Zinser}

%Andreas Albert, Mihailo Backovic, Antonio Boveia, Oliver Buchmueller, Giorgio Busoni, Albert De Roeck, Caterina Doglioni,  Tristan DuPree, Malcolm Fairbairn, Marie-Helene Genest, Stefania Gori, Giuliano Gustavino, Kristian Hahn, Ulrich Haisch, Philip C. Harris, Dan Hayden, Valerio Ippolito, Isabelle John, Felix Kahlhoefer, Suchita Kulkarni, Greg Landsberg, Steven Lowette, Kentarou Mawatari, Antonio Riotto, William Shepherd, Tim M.P. Tait, Emma Tolley, Patrick Tunney, Bryan Zaldivar, Markus Zinser

\hfill CERN-LPCC-2017-01

\abstract{
Weakly-coupled TeV-scale particles may mediate the interactions between normal matter and dark matter. If so, the LHC would produce dark matter through these mediators, leading to the familiar ``mono-$X$" search signatures, but the mediators would also produce signals without missing momentum via the same vertices involved in their production. This document from the LHC Dark Matter Working Group suggests how to compare searches for these two types of signals in case of vector and axial-vector mediators, based on a workshop that took place on September 19/20, 2016 and subsequent discussions. These suggestions include how to extend the spin-1 mediated simplified models already in widespread use to include lepton couplings. This document also provides analytic calculations of the relic density in the simplified models and reports an issue that arose when ATLAS and CMS first began to use preliminary numerical calculations of the dark matter relic density in these models.}
\maketitle

%%%%%%%%%%%%%%%%
\section{Introduction} 
\label{Intro}

During the LHC Run-2, ATLAS and CMS searches for dark matter (DM) using missing transverse energy signals have begun to use a common set of simplified models, reviewed by the ATLAS/CMS Dark Matter Forum~(DMF), to describe how DM would be produced~\cite{Abercrombie:2015wmb}. The models involve TeV-scale mediating particles that couple to quarks and a Dirac fermion DM candidate. The coupling to DM leads to collision events where a high-energy Standard Model (SM) final state recoils against invisible DM particles. Many types of accompanying SM particles are possible, often arising from initial-state radiation, creating a broad set of possible signals involving missing transverse energy (MET). The coupling to quarks, which permits the LHC to produce the mediating particles, also allows the mediators to decay to jets~\cite{Dreiner:2013vla,Chala:2015ama,Fairbairn:2016iuf} or possible to top-quark pairs~\cite{Chala:2015ama,ATLAS:2016pyq,Bauer:2017ota}. Such events, which lack substantial MET, could be used to fully or partially reconstruct the mass and other properties of the mediators.

The LHC Dark Matter Working Group (WG), established by ATLAS, CMS, and the LHC Physics Centre at CERN (LPCC) as the successor of the ATLAS/CMS DMF~\cite{Abercrombie:2015wmb}, has recommended a set of standardized plots for comparing MET searches channels that differ in the accompanying SM recoil~\cite{Boveia:2016mrp}. The recommendations include depicting the results of these searches in slices of DM mass versus mediator mass for fixed values of the mediator couplings to DM and SM particles. However the WG did not address how these comparisons could incorporate searches for fully-visible decays of the mediators.

As ATLAS and CMS adopted the recommendations for their Run-2 results, both produced preliminary comparisons between visible-decay and invisible-decay searches, starting with an ATLAS comparison of mono-jet, mono-photon, and di-jet searches in the DM-mediator mass plane for a single choice of couplings~\cite{ATLASsummaryplots}, followed by a  comparison of CMS results~\cite{CMS_SummaryPlots_ICHEP} that were also extrapolated to the DM-nucleon cross section for direct-detection DM searches (see also~\cite{Sirunyan:2016iap} for updated results on di-jet resonances). Both ATLAS and CMS results also depict values of the mass parameters where the simplified model reproduces the observed DM density in the standard thermal relic scenario.

The present document discusses some of what has been learned while preparing the above results and includes additional recommendations, stemming from the discussion at the public meeting of the WG in September 2016. Section~\ref{sec:models} adds couplings to leptons for the $s$-channel vector and axial-vector simplified models and provides additional benchmark coupling scenarios that illustrate the relationships amongst the various visible and invisible mediator searches. Section~\ref{sec:relic} discusses a deficiency in the relic density calculations commonly used for the first Run-2 results~\cite{Boveia:2016mrp,ATLASsummaryplots,CMS_SummaryPlots_ICHEP} and compares a new computation with version 2.0.6 of \maddm with the results of an analytic calculation. 

%%%%%%%%%%%%%%%%

%%%%%%%%%%%%%%%%
\section{Lepton couplings for simplified DM models}
\label{sec:models}

The simplified models recommended by the ATLAS/CMS DMF~\cite{Abercrombie:2015wmb} assume that DM is a Dirac fermion~$\chi$ and there is an additional heavy particle mediating the SM-DM interaction (the ``mediator"). In the most basic set of these models, the mediator is a vector, an axial-vector, a scalar or a pseudo-scalar boson. So far, ATLAS and CMS have focused on the subset of the models where the mediator is exchanged in the $s$-channel. These models contain four free parameters. In the vector and axial-vector models, the parameters are the DM mass~$\mDM$, the mediator mass~$\mmed$, the coupling~$\gDM$ of a mediator-DM-DM vertex, and the coupling~$\gq$ universal to all mediator-quark-quark vertices. In the scalar and pseudo-scalar models, a quark-mass-dependent Yukawa factor scales the coupling of the mediator-quark-quark vertices to avoid violating flavor constraints. These four quantities parameterize the production rate of the mediator in proton-proton collisions, its quark and DM decay rates, and the kinematic distributions of signal events.

Complete models of DM can contain mediators that may have (or require for consistency) couplings to other SM particles that are not found in the simplified models above. Such couplings would introduce additional decay modes of the mediator at the LHC as well as further DM annihilation channels in the relic density calculation. In this section, we discuss why and how to add lepton couplings to the vector and axial-vector simplified models, provide formulas for the total decay width of the mediators, and discuss the implementations of these models that are currently available. We then propose four benchmark scenarios for comparing di-jet, di-lepton, and mono-$X$ searches, based on rough estimates of the sensitivity of these searches with $30 \, {\rm fb}^{-1}$ of LHC data. We also comment on the interference between the mediator di-lepton process and the Drell-Yan backgrounds to di-lepton searches. We postpone the discussion of scalar and pseudo-scalar models to a future document. 

%%%%%%%%%%%%%%%%

\subsection{Charged lepton couplings in vector and axial-vector simplified models}
\label{sub:vecAxial}

Simplified models are designed to capture the coarse details of collider phenomenology found in complete, rigorously-derived theories of new physics, without the attendant complexity of the full theory, particularly physics at energy scales that cannot be accessed at the collider. The ATLAS/CMS DMF focused on the phenomenology of MET signatures at the LHC. In the simplified DMF models involving spin-1 mediators,\footnote{In this document, we will focus on the case of spin-1 mediators, and postpone the discussion of scalar and pseudo-scalar mediators to future work.} quark couplings provide $pp$ collider production, and DM couplings provide the decays to DM. These two couplings set a ``minimal width" for the spin-1 resonance. 

When adapting the simplified DMF models to the phenomenology of fully-visible signatures, one should more closely consider the effects of the additional% Uli, optional 
couplings. Among these, couplings to charged leptons are often found or even required in complete theories. They are sometimes necessary in order to construct a consistent theory, for example in minimal completions of the axial-vector model~~\cite{Kahlhoefer:2015bea,Jacques:2016dqz} or in models with extended Higgs sectors~\cite{Arcadi:2013qia,Bauer:2016gys}. They often appear in anomaly-free spin-1 mediator models~\cite{Carena:2004xs}, see also Section 3.3.2 of~\cite{Boveia:2016mrp}. They may also be induced through radiative corrections (e.g.~through quark loops that lead to $Z^\prime$--$Z$ mixing). The near-ubiquity of lepton couplings in full theories motivates including them when searching for visibly-decaying spin-1 mediators.

The DMF spin-1 simplified models can be easily extended with couplings to charged leptons $g_\ell$, equal for all lepton flavours. Assuming that the new interactions conserve parity, mediator vertices with leptons will have the same Lorentz structure as the vertices with quarks. We then obtain the following interaction Lagrangians for the vector and axial-vector $Z^\prime$ mediator models:
\begin{align}
\label{eq:AV1}
\mathcal{L}_{\text{vector}} &=- \gDM \, Z^\prime_{\mu} \, \bar{\chi}\gamma^{\mu}\chi -   \gq  \sum_{q=u,d,s,c,b,t} Z^\prime_{\mu} \, \bar{q}\gamma^{\mu}q  - g_\ell  \sum_{\ell=e,\mu,\tau} Z^\prime_{\mu} \, \bar{\ell}\gamma^{\mu}\ell\,, \\
\label{eq:AV2} 
\mathcal{L}_{\text{axial-vector}}&=- \gDM \, Z^\prime_{\mu} \, \bar{\chi}\gamma^{\mu}\gamma_5\chi - \gq \sum_{q=u,d,s,c,b,t} Z^\prime_{\mu} \, \bar{q}\gamma^{\mu}\gamma_5q
-  g_\ell  \sum_{\ell=e,\mu,\tau} Z^\prime_{\mu} \, \bar{\ell}\gamma^{\mu}\gamma_5\ell\,.
\end{align}
Notice that the generation-universality of the couplings $\gq$ and $g_\ell$ guarantees that these spin-1 models are consistent with --- but more restrictive than --- the minimal flavour violation (MFV) assumption~\cite{D'Ambrosio:2002ex}, imposed to evade constraints from flavor physics. 

Adding lepton couplings allows the mediator to decay to charged lepton pairs at tree level. For many values of $g_\ell$, this will lead to stringent bounds from searches for di-lepton resonances.

\subsection{Neutrino couplings in vector and axial-vector simplified models}

Following the reductionist philosophy of simplified models, the DMF did not build strict theoretical self-consistency into its models. For example, the simplified models do not specify how the $Z^\prime$ boson acquires a mass nor does the formulation of the models explicitly require gauge invariance. When adjusting the focus of the simplified models beyond mono-jet like searches to also include direct searches for the mediators, neglecting these aspects becomes less justified. While a discussion of mass generation in spin-1 simplified models is beyond the scope of this document, we will in the following explain how gauge invariance restricts the lepton couplings of the spin-1 models.

In the case at hand, gauge invariance requires a relation of the couplings of the spin-1 mediator to charged leptons and the  left-handed neutrinos. For both the vector and the axial-vector model, the Lagrangian that describes relevant neutrino interactions for each neutrino flavor takes the form:
\begin{align}
\label{eq:neu}
\mathcal{L}_\nu = - g_\nu \sum_{i=e,\mu,\tau} Z^\prime_\mu \bar \nu_i \gamma^\mu \frac{1}{2}(1-\gamma_5) \nu_i \, .
\end{align}
The relation required between $g_\nu$ and $g_\ell$ differs in the two models. For the vector model, $g_\nu = g_\ell$, whereas for the axial-vector model, $g_\nu = - g_\ell$. Because right-handed neutrinos are absent in the SM, the coupling of the mediator to neutrinos necessarily breaks parity and therefore has a different Lorentz structure from the coupling to charged leptons.\footnote{Because of the parity violation, it is strictly speaking no longer correct to distinguish between the vector and the axial-vector model for the neutrino sector. Nevertheless, we will continue to use these terms as long as parity is a symmetry of the interactions of quarks and DM.}

The new coupling $g_\nu$, implied by gauge invariance, has an important consequence for the phenomenology of MET searches: it supplies an additional invisible decay channel, which may enhance certain mono-$X$ signals. 

\subsection{Width formulas and model implementation}

Including leptonic couplings the partial decay widths of the vector mediator are given by
\begin{align}
\Gamma_{\text{vector}}^{\chi\bar{\chi}} & = \frac{\gDM^2 \hspace{0.25mm} \mmed}{12\pi} 
 \left (1-4 \hspace{0.25mm}  z_{\rm{DM}} \right )^{1/2} \left(1 + 2 \hspace{0.25mm}  z_{\rm{DM}} \right) \, , \\
\Gamma_{\text{vector}}^{q\bar{q}} & = \frac{\gq^2 \hspace{0.25mm}  \mmed}{4\pi} 
 \left ( 1-4 \hspace{0.25mm}  z_q \right )^{1/2}   \left(1 + 2 \hspace{0.25mm}  z_q \right) \, , \\
 \Gamma_{\text{vector}}^{\ell\bar{\ell}} & = \frac{g_\ell^2 \hspace{0.25mm}  \mmed}{12\pi} 
 \left ( 1-4 \hspace{0.25mm}  z_\ell \right )^{1/2}   \left(1 + 2 \hspace{0.25mm}  z_\ell \right) \, , \\
 \Gamma_{\rm vector}^{\nu\bar{\nu}} & = \frac{g_\ell^2}{24\pi} \mmed \, , 
\end{align}
where $z_i = m_i^2/\mmed^2$ with $i={\rm DM},q,\ell$, and the three different types of contributions to the decay width vanish for $\mmed < 2 \hspace{0.25mm}  m_i$. 
The corresponding expressions for the axial-vector mediator are
\begin{align}
\Gamma_{\text{axial-vector}}^{\chi\bar{\chi}} & = \frac{\gDM^2 \, \mmed}{12\pi} 
\left ( 1-4 \hspace{0.25mm} z_{\rm{DM}} \right ) ^{3/2} \,, \\ 
  \Gamma_{\text{axial-vector}}^{q\bar{q}} & =  \frac{\gq^2 \, \mmed}{4\pi} 
\left ( 1-4 \hspace{0.25mm} z_q \right ) ^{3/2} \, , \\
  \Gamma_{\text{axial-vector}}^{\ell\bar{\ell}} & =  \frac{g_\ell^2 \, \mmed}{12\pi} 
\left ( 1-4 \hspace{0.25mm} z_\ell \right ) ^{3/2} \, , \\
 \Gamma_{\text{axial-vector}}^{\nu\bar{\nu}} & = \frac{g_\ell^2}{24\pi} \mmed \, .   
\end{align}

Chapter 4 of the ATLAS/CMS DMF report~\cite{Abercrombie:2015wmb} provides guidelines for simulating the models it discusses, along with a reference implementation~\cite{LHCDMFmodels}. Another more recent implementation of the spin-1 DMF models that provides next-to-leading order plus parton shower accuracy in the {\sc MadGraph5\_aMC\@NLO} framework~\cite{Alwall:2014hca} has been presented in~\cite{Backovic:2015soa}. The corresponding {\sc UFO} file \cite{Degrande:2011ua} has been obtained with {\sc FeynRules~2}~\cite{Alloul:2013bka} and can be found at \cite{DMsimp}. The original implementation has been modified to include the lepton couplings discussed above. 

\subsubsection{Benchmark scenarios for simplified models with lepton couplings}

In an earlier document~\cite{Boveia:2016mrp}, this WG recommended a set of standardized plots for comparing results from different MET search channels in these models, including depicting the search results in slices of DM mass versus mediator mass for fixed values of the mediator couplings to DM and SM particles. Because in the spin-1 case the differences in the various signals arise from initial state radiation, their rates relative to one another are fixed by SM couplings, not the new couplings entering  the simplified model. When using the same plots for subsequent comparisons with searches for fully-visible signatures, whose signal rates relative to the invisible channels do depend on the couplings in the simplified model, albeit in straightforward ways, it becomes crucial to convey how the relative strength of each search varies with the choice of couplings.

To solve this problem, the strategy employed by ATLAS and CMS has been to show slices of DM mass versus mediator mass for one or more sets of ``benchmark" coupling values that illustrate the complementary strengths of the different searches, as in~\cite{Boveia:2016mrp,ATLASsummaryplots,CMS_SummaryPlots_ICHEP}. When introducing lepton couplings, we recommend the following four scenarios with different relative sizes of quark and lepton couplings:
\begin{itemize}
    \item V1: Vector model with couplings only to quarks: $\gdm=1.0$, $\gq = 0.25$, $g_\ell = 0$.
    \item V2: Vector model with a small couplings to leptons: $\gdm=1.0$, $\gq = 0.1$, $g_\ell = 0.01$.
    \item A1: Axial-vector model with couplings only to quarks: $\gdm=1.0$, $\gq = 0.25$, $g_\ell = 0$.
    \item A2: Axial-vector model with equal couplings to quark and leptons: $\gdm=1.0$, $g_q=g_\ell=0.1$.
\end{itemize}

Scenarios V1 and A1 are the simplified models already in use. Scenario A2 represents a representative case found in the simplest complete models with axial-vector $Z^\prime$ bosons~\cite{Kahlhoefer:2015bea}, 
and illustrates the typical impact of searches for di-lepton resonances in these models. When the mediator is a pure vector, however, one can find $g_\ell \ll g_q$. This is for example the case if the mediator couples only to quarks (and DM) at tree-level and  obtains couplings to leptons only from mixing with the neutral SM gauge bosons at loop-level. In such a scenario one naturally expects $g_\ell/g_q = {\cal O} (0.1)$~\cite{Duerr:2016tmh} with the precise value of the ratio depending on the exact model realisation. 
Scenario V2 provides a benchmark for this plausible but more pessimistic (from the di-lepton point of view) possibility. The specific value, $g_\ell = 0.1 g_q$, is chosen so that searches for di-jet and di-lepton resonances will have comparable sensitivity. The contribution to the signal width from neutrino couplings is negligible in both scenarios A2 and V2, and it can be ignored. 

Because LHC searches become sensitive to smaller production cross sections as data are collected, it is also meaningful to consider smaller values of $g_q$ (and hence $g_\ell$) with respect to the initial Run-2 benchmarks. For $30 \, {\rm fb}^{-1}$ of data, we recommend $\gq = 0.1$ (and $\gDM = 1$). For this smaller quark coupling, with $g_\ell = 0.1$ for Scenario A2 and $g_\ell = 0.01$ for V2, the total decay width of the mediator is up to 3.2\%. 

To consider broader mediator widths while at the same time further suppressing constraints from searches for resonant two-body decays, it may also be interesting to consider larger values of $\gDM$. For example, the spin-1 models are still well within the perturbative regime for $\gDM = 2$, predicting a mediator width of only 6\% (for $\gq = g_\ell = 0.1$).

\subsection{Interference effects in di-lepton searches}

Both the ATLAS and the CMS collaboration have already conducted detailed searches with Run-1~and~Run-2 data for massive di-lepton resonances, using assorted spin-1 models~\cite{Aad:2014cka,Khachatryan:2014fba,Aaboud:2016cth,Khachatryan:2016zqb,Khachatryan:2016qkc,ATLAS-CONF-2016-045}. These searches have concentrated on narrow resonant signals in the di-lepton invariant mass spectrum $d \sigma/d m_{\ell \ell}$, where one can ignore interference effects between the signal and the SM Drell-Yan background. Such interference 
effects cannot be neglected in these searches if 
they significantly modify the size of the signal or distort its shape. 

To assess the size of interference effects for the four benchmark scenarios introduced in the previous section, we have recalculated  $d \sigma/d m_{\ell \ell}$ before and after taking the interference into account. The benchmark model with the largest relative width $\Gamma_{\rm med}/M_{\rm med}$ is 
scenario~A2 with $g_q=g_\ell =0.1$. Setting $g_{\rm DM} = 2$ to exacerbate the effects of width in this scenario, we still find that the interference effects never exceed $5\%$ when $d \sigma/d m_{\ell \ell}$ is integrated between $m_{\ell \ell} \in [M_{\rm med}-5 \Gamma_{\rm med}, M_{\rm med}+5 \Gamma_{\rm med}]$. The same conclusion holds when the di-lepton pairs are required to pass the selections imposed in the ATLAS and CMS dilepton searches~\cite{Aaboud:2016cth,Khachatryan:2016zqb}. Therefore we suggest such effects can be neglected when setting limits on the parameter space of spin-1 $s$-channel simplified models.\footnote{Starting from version 2.0 of the {\sc DMsimp} simplified model implementation~\cite{DMsimp}, interference effects in di-lepton resonance searches can be calculated for spin-1 $s$-channel simplified models.} We find worth mentioning, however, that for simplified models 
spin-0 $s$-channel mediators, interference effects are instead relevant in $t\bar{t}$ searches~\cite{Dicus:1994bm,Frederix:2007gi,Djouadi:2015jea,Craig:2015jba,Jung:2015gta,Bernreuther:2015fts,Gori:2016zto,Carena:2016npr,ATLAS:2016pyq,Bauer:2017ota}.

%%%%%%%%%%%%%%%%
\section{Relic Density}
\label{sec:relic}

In the standard thermal relic ``freeze-out" picture of the early Universe,
the annihilation rate of DM particles into normal matter determines the temperature at which DM decouples from thermal equilibrium and sets the DM density observed today.
One can use the simplified models discussed in Section~\ref{sec:models} to predict the relic density and compare with measurements such as the most recent results by the Planck collaboration~\cite{Ade:2015xua} to gain insight on interesting regions of the model parameter space. In order to do so, one has to make the following assumptions:
\begin{itemize}
\item The DM annihilation cross section receives only contributions from the interactions of the simplified model, while possible additional degrees of freedom and couplings not included in the model are irrelevant. 
\item The DM number density in the Universe today is entirely determined by  the DM annihilation cross section predicted by the simplified model. In particular, no additional mechanisms exist that enhance or deplete the relic density.
\end{itemize}

It it important to realize that if one or both of these assumptions are violated there is no strict correlation between the relic density and the strength of mono-$X$ signals. For instance, if DM is overproduced, the relic density can be reduced if the DM has large annihilation cross sections to new hidden sector states. These states might however not be directly accessible at LHC energies.  Conversely, the correct DM relic density can still be obtained  if the DM is underproduced. For instance, if the hidden sector carries an particle-antiparticle asymmetry (similar to the baryon asymmetry) then this necessarily leads to a larger relic density compared to the conventional freeze-out picture. 

In this section, we assume that the two aforementioned assumptions are satisfied, and  present an analytic calculation of the relic density for the dominant annihilation processes that involve spin-0 and spin-1 mediators. We then provide numerical computations of the relic density for the  scalar, pseudo-scalar, vector, and axial-vector simplified model scenarios using version 2.0 of the \dmsimp implementation~\cite{DMsimp}
and version 2.0.6 of \maddm~\cite{Backovic:2013dpa,Backovic:2015tpt}. The Lagrangians for these models can be found in~\cite{Boveia:2016mrp} and references therein. 

For concreteness  the coupling values recommended for the first Run-2 results by the ATLAS/CMS DMF are used in this section. The couplings of the spin-1 mediator (vector or axial-vector) to SM quarks is chosen to be $g_q = 0.25$ and the lepton coupling value is set to zero, corresponding to scenarios V1 and A1 of Section~\ref{sub:vecAxial}. The coupling value of the spin-0 mediator (scalar or pseudo-scalar) to quarks is chosen to be $g_q=1.0$ with an implicit Yukawa scaling for all SM quarks. For both models, the coupling value of the mediator to DM particles is fixed to be $g_{\rm DM}=1$.
A complete set of relic density curves can be found in the LHC DM WG repository~\cite{RelicRepo}.

\subsection{Analytic expressions for the DM relic density}
\label{analyticrelic}

\begin{center}
\begin{figure}[!t]
\centering
\includegraphics[width=0.78\textwidth]{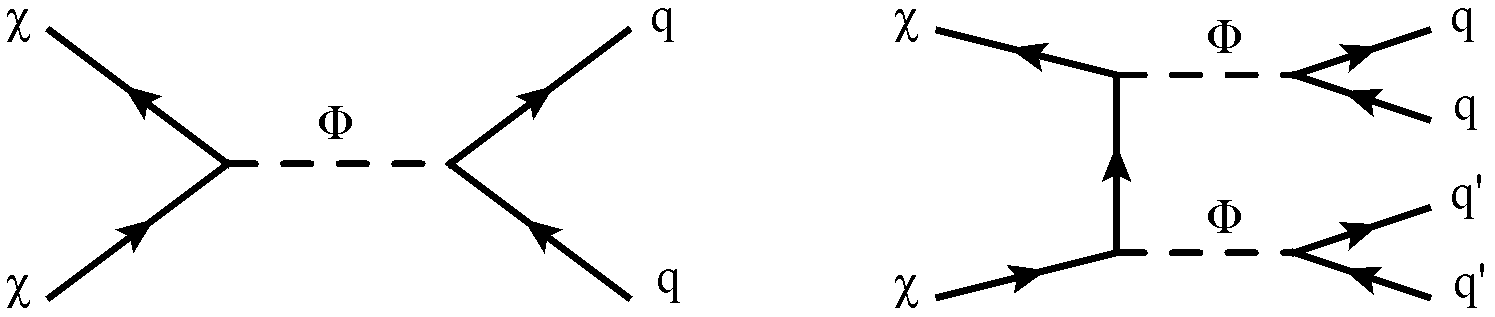} 
\vspace{4mm}
\caption{Feynman graphs of 
DM $s$-channel annihilation to quarks (left) and $t$-channel annihilation to a pair of mediators subsequently decaying to quarks (right). The exchanged~$\Phi$ particle(s) can be either (pseudo-)scalar or (axial-)vector mediator(s). 
}
\label{fig:relicprod}
\end{figure}
\end{center}

Figure~\ref{fig:relicprod} shows the two  
types of Feynman diagrams that are 
most important in the calculation of the relic density in the simplified models. The graph on the left-hand side illustrates DM annihilation through a single mediator in the $s$-channel, while the diagram on the right corresponds to DM annihilation to pairs of mediators via the $t$-channel.
For $M_{\rm med}/2 > m_{\rm DM}$, the $s$-channel process dominates, while the $t$-channel process gives the main contribution when the mediators can go on-shell, that is for $M_{\rm med} < m_{\rm DM}$. For some choices of mediator, e.g.~the pseudo-scalar simplified model, higher
order processes such as annihilation into three or
more mediators are also important if they are kinematically accessible \cite{Abdullah:2014lla}.

Analytic expressions for the annihilation cross sections can be derived separately for both Feynman diagrams in Figure~\ref{fig:relicprod}. 
In the case of the $s$-channel graphs we obtain 
\begin{align}
\sigma_{{\rm ann},s}^S \cdot v & = \sum_{q} \frac{N_c^q \hspace{0.125mm} g_{\rm DM}^2  \hspace{0.5mm} y_q^2 \hspace{0.25mm} g_q^2 \hspace{0.5mm}  \beta_q}{16 \pi}  \, \frac{m_{\rm DM}^2 - m_q^2}{\left (M_{\rm med}^2 - 4 m_{\rm DM}^2 \right )^2 + M_{\rm med}^2 \Gamma_{\rm med}^2} \; v^2 \,, \label{eq31} \\
\sigma_{{\rm ann},s}^P \cdot v & = \sum_{q} \frac{N_c^q \hspace{0.125mm} g_{\rm DM}^2 \hspace{0.5mm} y_q^2 \hspace{0.25mm} g_q^2  \hspace{0.5mm} \beta_q}{4 \pi}  \, \frac{m_{\rm DM}^2}{\left (M_{\rm med}^2 - 4 m_{\rm DM}^2 \right )^2 + M_{\rm med}^2 \Gamma_{\rm med}^2} \,, \label{eq32} \\
\sigma_{{\rm ann},s}^V \cdot v & = \sum_{q} \frac{N_c^q \hspace{0.125mm} g_{\rm DM}^2 \hspace{0.25mm} g_q^2 \hspace{0.5mm} \beta_q}{2 \pi} \,  \frac{2 m_{\rm DM}^2+ m_q^2}{\left (M_{\rm med}^2 - 4 m_{\rm DM}^2 \right )^2 + M_{\rm med}^2 \Gamma_{\rm med}^2} \,, \label{eq33} \\
\sigma_{{\rm ann},s}^A \cdot v & = \sum_{q} \frac{N_c^q \hspace{0.125mm} g_{\rm DM}^2 \hspace{0.25mm} g_q^2  \hspace{0.5mm} \beta_q}{2 \pi}  \frac{ m_{q}^2 \left ( 4 m_{\rm DM}^2- M_{\rm med}^2  \right)^2}{M_{\rm Med}^4 \left [ \left (M_{\rm med}^2 - 4 m_{\rm DM}^2 \right )^2 + M_{\rm med}^2 \Gamma_{\rm med}^2 \right ] } \,, \label{eq34}
\end{align}
where the sum includes all quarks with $m_q \leq m_{\rm DM}$, $N_c^q =3$, $\beta_q = \sqrt{1 - m_q^2/m_{\rm DM}^2}$ and~$v$ is the 
relative velocity of the DM pair. Notice that in the pseudo-scalar, vector, and axial-vector case the $s$-channel annihilation cross section proceeds via $s$-wave,~i.e.~it is of ${\cal O} (v^0)$, while in the case of scalar exchanges DM annihilation is $p$-wave suppressed,~i.e.~it is of ${\cal O} (v^2)$. The corresponding annihilation cross sections into charged leptons can be obtained from the above expressions by a suitable replacement of color factors and SM fermion masses. 

In the case of the $t$-channel diagrams we instead find  the following annihilation cross sections
\begin{align}
\sigma_{{\rm ann},t}^S \cdot v & =  \frac{ g_{\rm DM}^4   \hspace{0.5mm} \beta_{\rm med}}{24 \pi}  \, \frac{ m_{\rm DM}^2 \left ( 9 m_{\rm DM}^4  - 8 m_{\rm DM}^2 M_{\rm med}^2  + 2 M_{\rm med}^4   \right )}{\left (M_{\rm med}^2 - 2 m_{\rm DM}^2 \right )^4}  \; v^2 \,, \label{eq35} \\
\sigma_{{\rm ann},t}^P \cdot v & =  \frac{ g_{\rm DM}^4   \hspace{0.5mm} \beta_{\rm med}}{24 \pi} \,  \frac{m_{\rm DM}^2 \left (  m_{\rm DM}^2  - M_{\rm Med}^2  \right )^2}{\left (M_{\rm med}^2 - 2 m_{\rm DM}^2 \right )^4}  \; v^2 \,, \label{eq36} \\
\sigma_{{\rm ann},t}^V \cdot v & =  \frac{ g_{\rm DM}^4   \hspace{0.5mm} \beta_{\rm med}}{4 \pi}  \, \frac{m_{\rm DM}^2 - M_{\rm med}^2 }{\left (M_{\rm med}^2 - 2 m_{\rm DM}^2 \right )^2}  \,, \label{eq37} \\
\sigma_{{\rm ann},t}^A \cdot v & =  \frac{ g_{\rm DM}^4   \hspace{0.5mm} \beta_{\rm med}}{4 \pi}  \, \frac{m_{\rm DM}^2 - M_{\rm med}^2  }{\left (M_{\rm med}^2 - 2 m_{\rm DM}^2 \right )^2} \,, \label{eq38}
\end{align}
for $M_{\rm med} \leq m_{\rm DM}$. Here $\beta_{\rm med} = \sqrt{1-M_{\rm med}^2/m_{\rm DM}^2}$ and one observes that the  annihilation of DM into a pair of mediators is $p$-wave ($s$-wave) in the case of spin-0 (spin-1) exchanges. From the above results it follows that in the case of scalar, vector, and axial-vector interactions both $s$-channel and $t$-channel annihilation has to be considered, while in the case of a pseudo-scalar typically only the $s$-channel contribution is relevant. We add that in some of the cases with non-vanishing $s$-wave contribution the $p$-wave contribution is nevertheless numerically relevant. This is for instance the case for  the axial-vector mediator where the $s$-wave contribution to the $s$-channel is helicity suppressed, while the $t$-channel receives only contributions from longitudinal polarizations. We do not provide the corresponding expressions here but included them in the numerical results presented below.

Using the velocity expansion\footnote{This expansion breaks down close to an $s$-channel resonance, 
making a numerical solution indispensable.}  $\sigma_{\rm ann} \cdot v = a + b v^2 + {\cal O} (v^4)$ the DM relic density  after freeze-out is approximately given by
\begin{equation}
    \Omega h^2 \simeq 0.12 \; \frac{1.6 \cdot 10^{-10} \, x_f \, {\rm GeV}^2}{a + \frac{3b}{x_f}} \,,
\end{equation}
where $x_f = m_{\rm DM}/T_f$ with $T_f$ the freeze-out temperature. In our comparison between analytic and numerical results we will employ $x_f = 28$. For this value of $x_f$ the correct relic abundance thus occurs in the ballpark of
\begin{equation}
2.2 \cdot 10^{-26} \, {\rm cm}^3/{\rm s} \simeq 4.5 \cdot 10^{-9} \, {\rm GeV}^{-2} \simeq a + 0.1 \hspace{0.25mm} b \,.
\end{equation}

\subsection{Numerical results}
\label{numericrelic}

One can improve upon the analytic calculation described above by performing a numerical calculation 
that also takes into account the thermal evolution of the Universe. The results presented in this subsection rely on \maddm version 2.0.6.  The \maddm package considers all tree-level $2\rightarrow2$ interactions between DM and SM particles. The processes are thermally averaged and the resulting relic density is computed.
Since \maddm does not yet automatically calculate the mediator width from the model parameters, the \dmsimp model was modified to use the mediator width formulas presented for instance in~\cite{Abercrombie:2015wmb,Boveia:2016mrp}. 
The DM density calculations provided in the previous LHC DM WG recommendations~\cite{Boveia:2016mrp} 
used an earlier version of \maddm which did not include  $t$-channel annihilation to pairs of mediators. Below we will comment on the effects that this omission has.  

\begin{center}
\begin{figure}[h]
\includegraphics[width=0.49\textwidth]{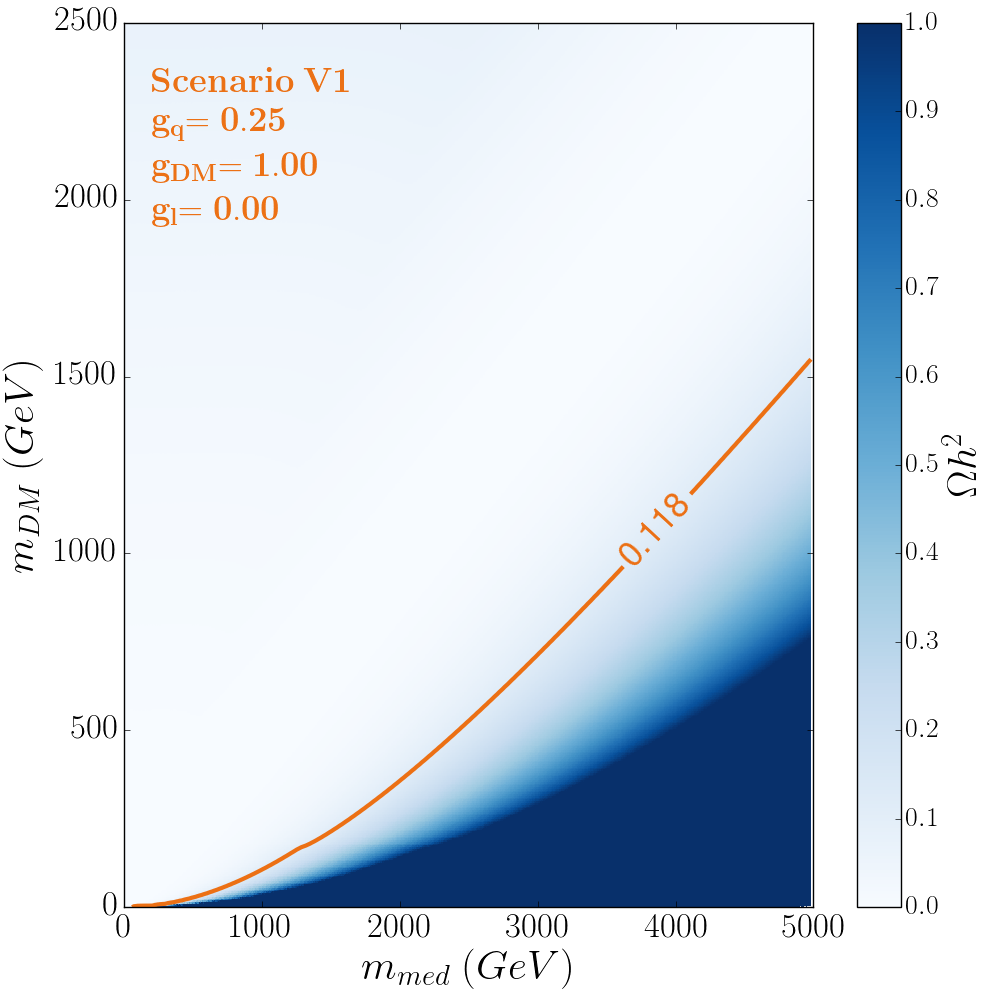} 
\includegraphics[width=0.49\textwidth]{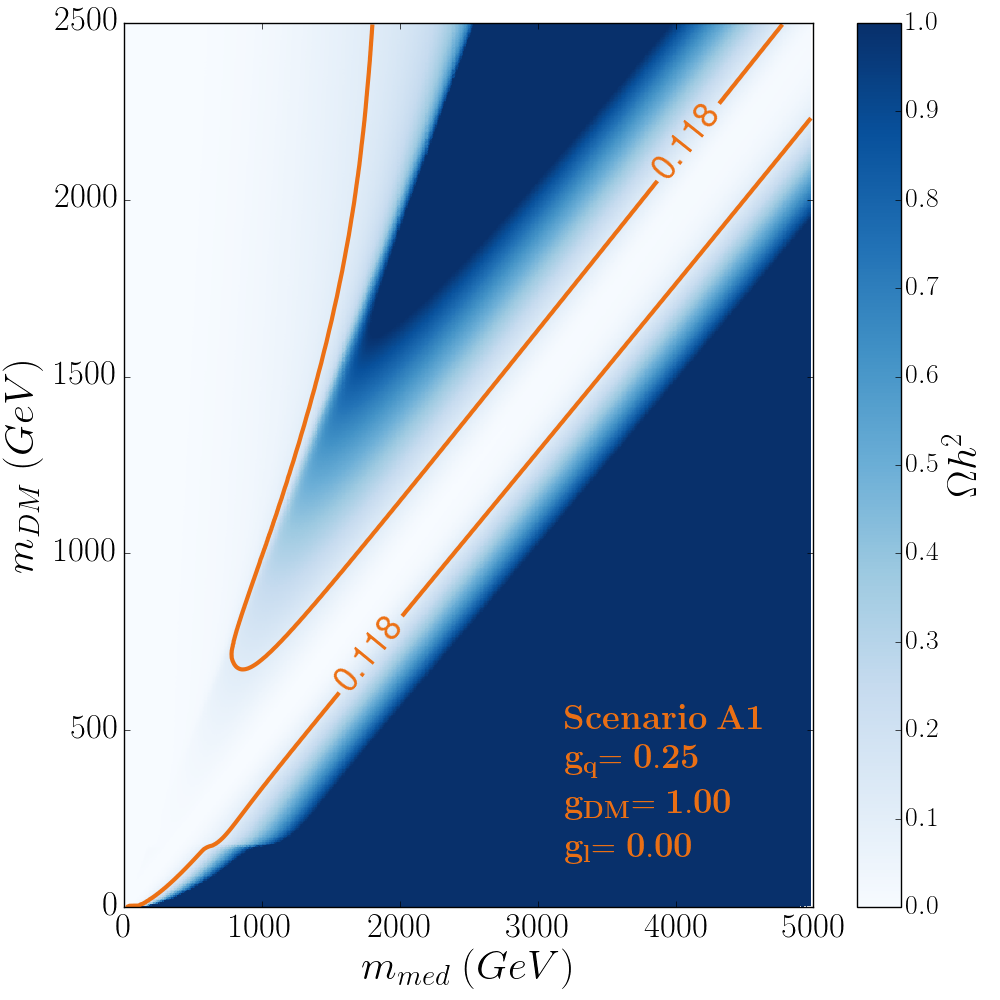} \\[2mm]
\includegraphics[width=0.49\textwidth]{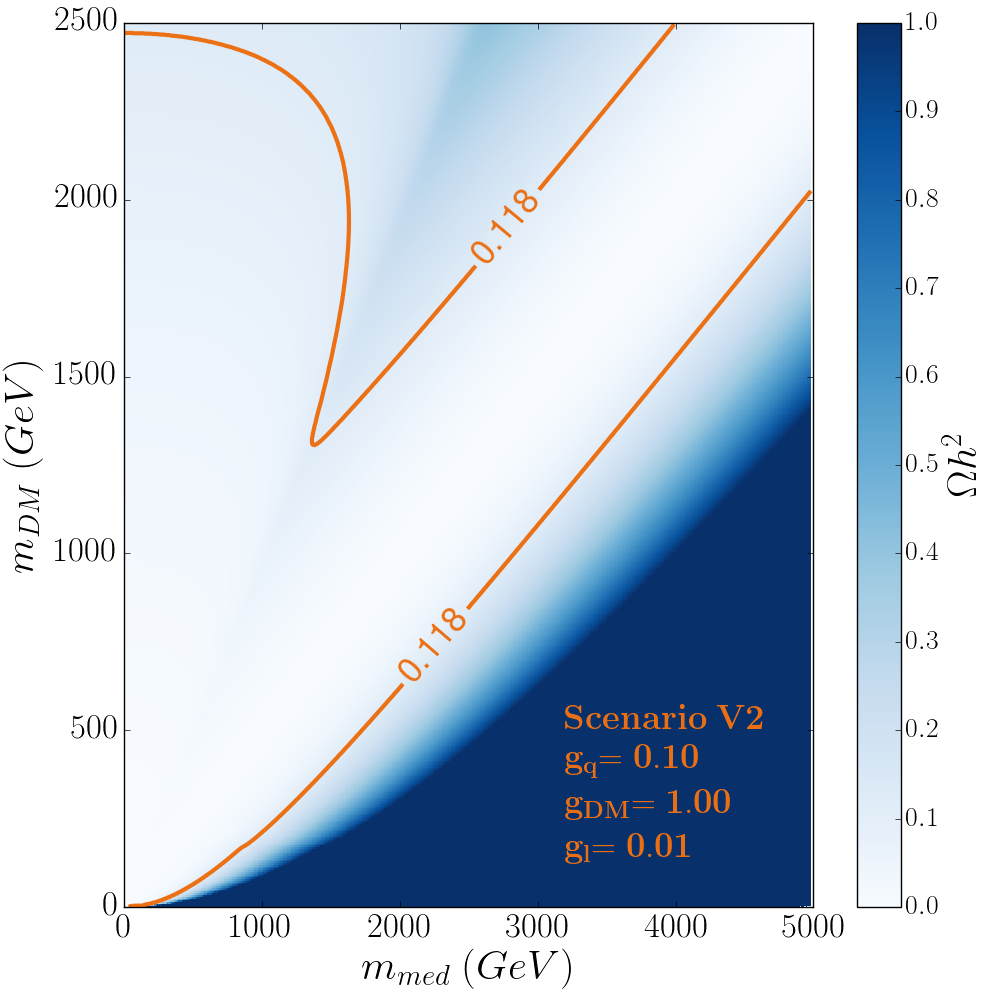} 
\includegraphics[width=0.49\textwidth]{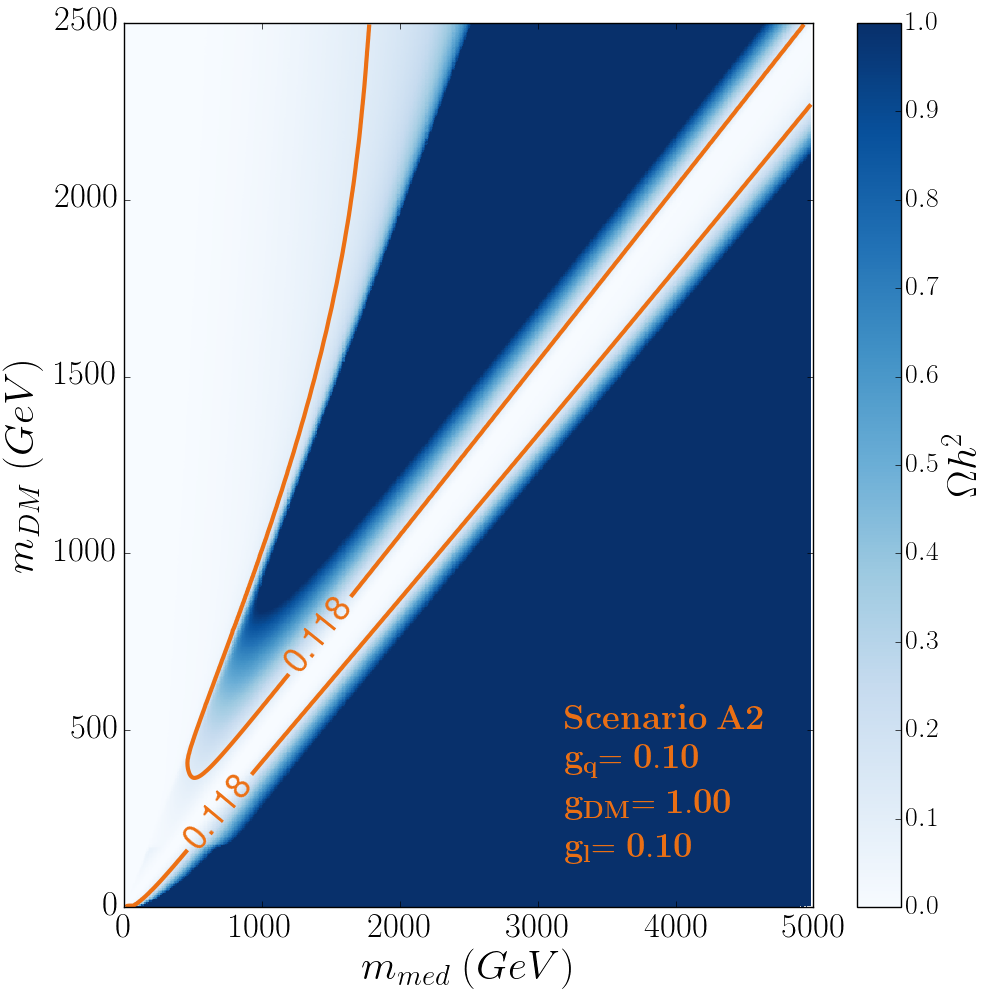} \\[-4mm]
\caption{The relic density $\Omega h^{2}$ in the $M_{\rm med}$--$m_{\rm DM}$ plane predicted by the four spin-1 scenarios  described in Section~\ref{sub:vecAxial}. All couplings to DM are set to unity ($g_{\rm DM}=1.0$). Top left: scenario V1, vector mediator with couplings only to quarks ($g_{ q}=0.25$). Top right:
scenario A1, axial-vector mediator with couplings only to quarks ($g_{q}=0.25$). 
 Bottom left: scenario V2, vector mediator with couplings to quarks and small couplings to leptons ($g_{ q}=0.1$ and $g_{\ell}=0.01$). Bottom right: scenario A2, axial-vector mediator with equal couplings to quarks and leptons ($g_{q}=0.1$ and $g_{\ell}=0.1$).
The contour lines correspond to the region of parameter space where the relic density is consistent with the value $\Omega h^2=0.118$.
}
\label{fig:DMBoundsScenarios}
\end{figure}
\end{center}

\begin{center}
\begin{figure}[t!]
\includegraphics[width=0.49\textwidth]{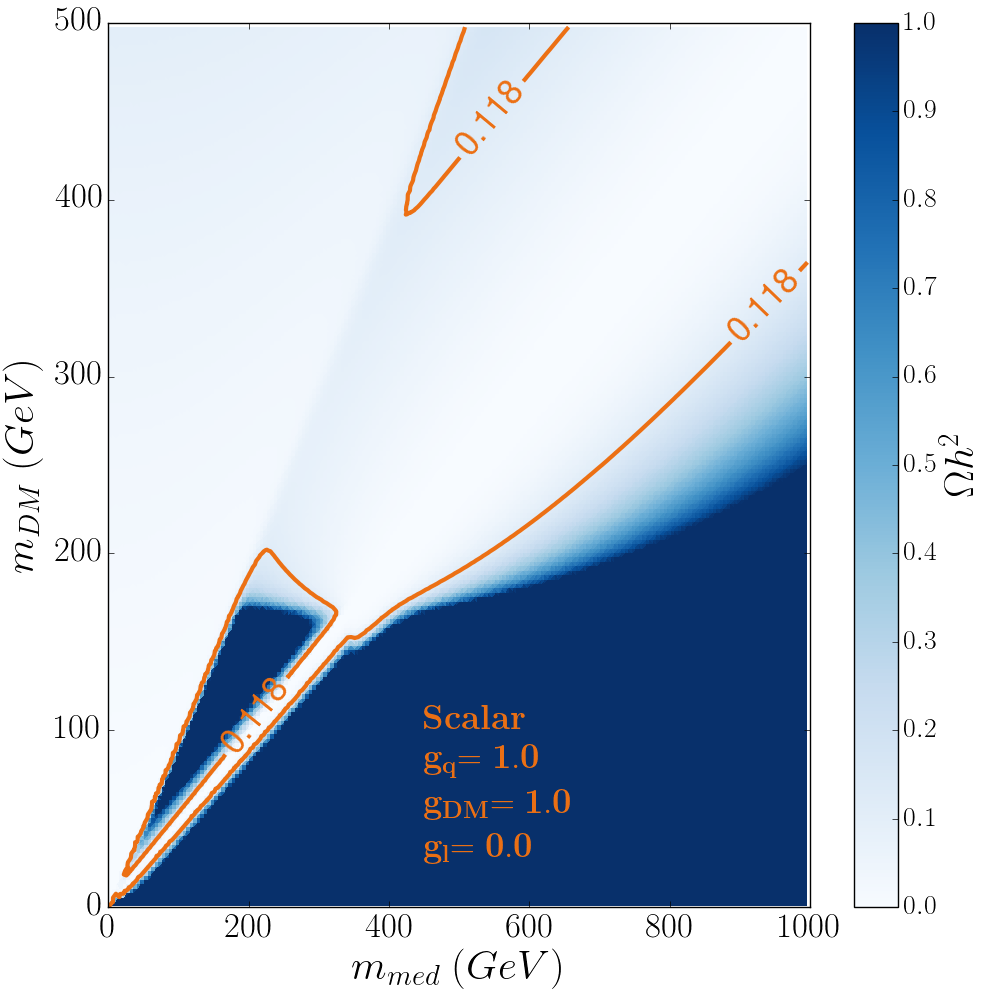} 
\includegraphics[width=0.49\textwidth]{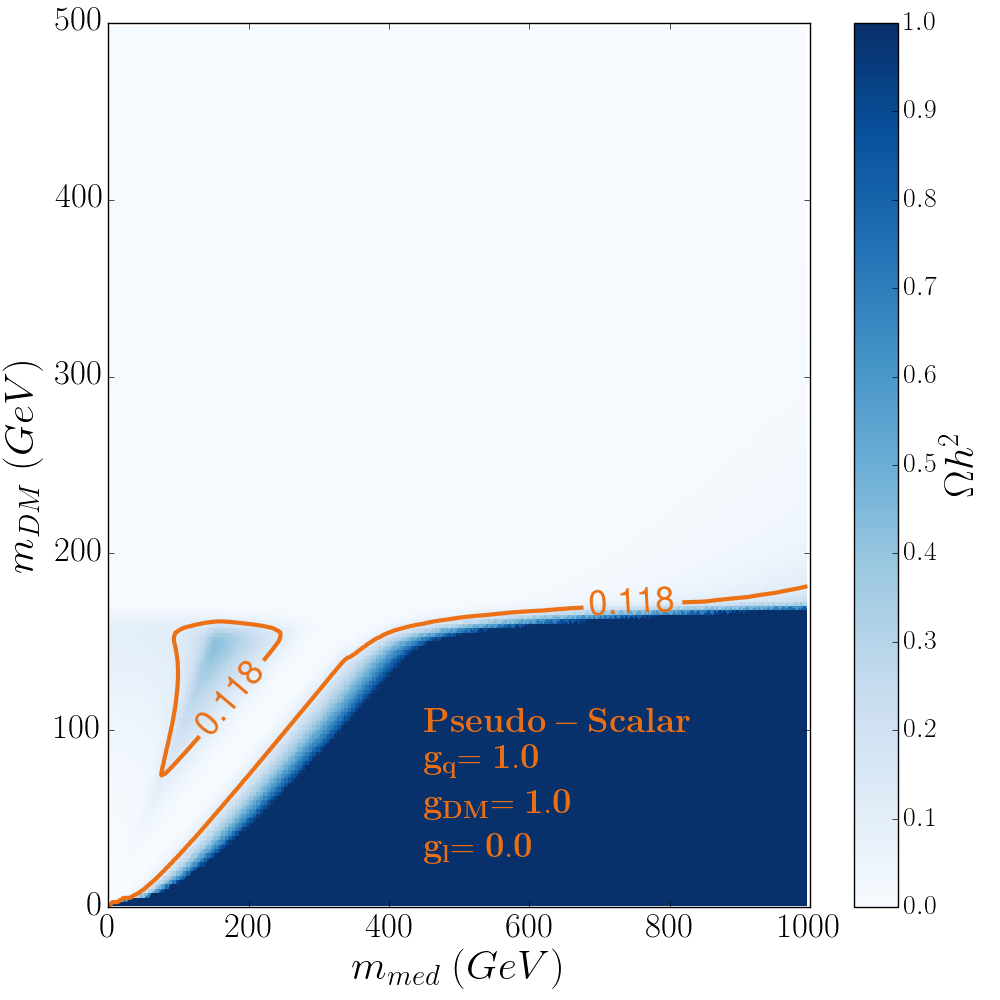} \\
\caption{The relic density $\Omega h^{2}$ in the $M_{\rm med}$--$m_{\rm DM}$ plane predicted by the scalar (left panel) and pseudo-scalar (right panel) mediator models with couplings $g_{\rm DM} = g_q =1.0$.
The contour lines indicate the observed value of the DM relic density $\Omega h^2=0.118$. }
\label{fig:DMBoundScalars}
\end{figure}
\end{center}

The panels in Figures~\ref{fig:DMBoundsScenarios} and \ref{fig:DMBoundScalars} show the predictions for the relic density $\Omega h^2$ in the $M_{\rm med}$--$m_{\rm DM}$ plane for spin-1 and spin-0 mediators, respectively. In the spin-1 case the  coupling scenarios described in Section~\ref{sub:vecAxial} are employed, while for the spin-0 models the standard coupling values $g_{\rm DM} = g_q =1$ and $g_{\ell} = 0$ have been used. The solid contours in all panels indicate the combination of masses for which the correct DM abundance $\Omega h^2 = 0.118$~\cite{Ade:2015xua} is obtained. The parts in the  $M_{\rm med}$--$m_{\rm DM}$ plane where the relic density is either higher or lower than the observed value are referred to as overabundant and underabundant regions, respectively.

One observes that all models predict an overabundance of DM for $M_{\rm med} \gg m_{\rm DM}$. While the shape and exact size of this region depend on the specific model realisation, larger quark couplings $g_{q}$ in general allow DM to annihilate into SM particles more efficiently, which reduces the parameter space over which overabundance can occur.

For the vector  scenario V1 (top left panel in Figure \ref{fig:DMBoundsScenarios}) only a single overabundant region with $M_{\rm med} \gg m_{\rm DM}$ is present. For the shown part of the $M_{\rm med}$--$m_{\rm DM}$ plane this case is fully consistent with previous results (see e.g.~\cite{Boveia:2016mrp,Busoni:2014gta, Pree:2016hwc}). In the axial-vector scenario~A1~(top right panel in Figure \ref{fig:DMBoundsScenarios}) the  overabundance region extends to higher $m_{\rm DM}$ values than in scenario V1. Additionally, there is an overabundance region above the diagonal $m_{\rm DM}=M_{\rm med}/2$. While this region is also present in the corresponding figures  of~\cite{Pree:2016hwc} its width in mediator mass is significantly narrower. 
The observed difference is due to $t$-channel annihilation diagrams to pairs of mediators that have not been included in the latter work but are relevant if  $M_{\rm med}<m_{\rm DM}$.

In both the vector scenario V2 and axial-vector scenario A2 (lower left and right plot in Figure~\ref{fig:DMBoundsScenarios}) the relic density is  
enhanced with respect to the corresponding scenarios V1~and~A1. This is a result of the quark couplings being smaller in  V2 and~A2 than in V1 and~A1. Decreasing the quark couplings however reduces the annihilation cross section, which in turn  leads to an overabundance of DM for larger parts of the $M_{\rm med}$--$m_{\rm DM}$ plane. We add that for scenario V2 with  $g_{\ell} = 0.01$, DM annihilation into leptons has essentially no effect on $\Omega h^2$.  In scenario A2 with $g_{\ell} = 0.1$ the relic density is instead slightly suppressed in the whole $M_{\rm med}$--$m_{\rm DM}$ plane compared to a model with quark couplings only.  

Notice finally that in the case of axial-vector mediation, $s$-channel annihilation proceeds via $s$-wave but is helicity suppressed, while for vector mediators no such suppression occurs $\big($cf.~(\ref{eq33}) and (\ref{eq34})$\big)$. This feature qualitatively explains why the regions with DM overabundance are typically larger for axial-vector scenarios than for vector models. 

For the scalar simplified model (left panel in Figure~\ref{fig:DMBoundScalars}), the overabundance region for small $m_{\rm DM}$ is fades out for $m_{\rm DM}$ values above the top threshold $m_t$, above which annihilation of DM pairs into top-quark pairs is allowed. Additional overabundance regions occur for $M_{\rm med} > m_{\rm DM} > M_{\rm med}/2$, where the upper bound is due to the onset of mediator pair production and the lower bound reflects the resonant enhancement of DM annihilation to SM particle pairs. A region of overabundance at $M_{\rm med} > m_{\rm DM}$ in the predictions shown in~\cite{Pree:2016hwc} is now underabundant after including the $t$-channel annihilation contributions.

The pseudo-scalar simplified model (right panel in Figure~\ref{fig:DMBoundScalars}) is similar to the scalar scenario, with less pronounced regions of overabundance due to the increased annihilation cross section from $s$-channel $s$-wave contributions. Consequently, no regions of overabundance are observed above the top threshold and the triangular region present for the scalar model at $M_{\rm med}/2 < m_{\rm DM}$ is reduced in size compared to~\cite{Pree:2016hwc}. 

\subsection{Comparison of analytic results to full numerical calculations}

\begin{center}
\begin{figure}[t!]
\includegraphics[width=0.49\textwidth]{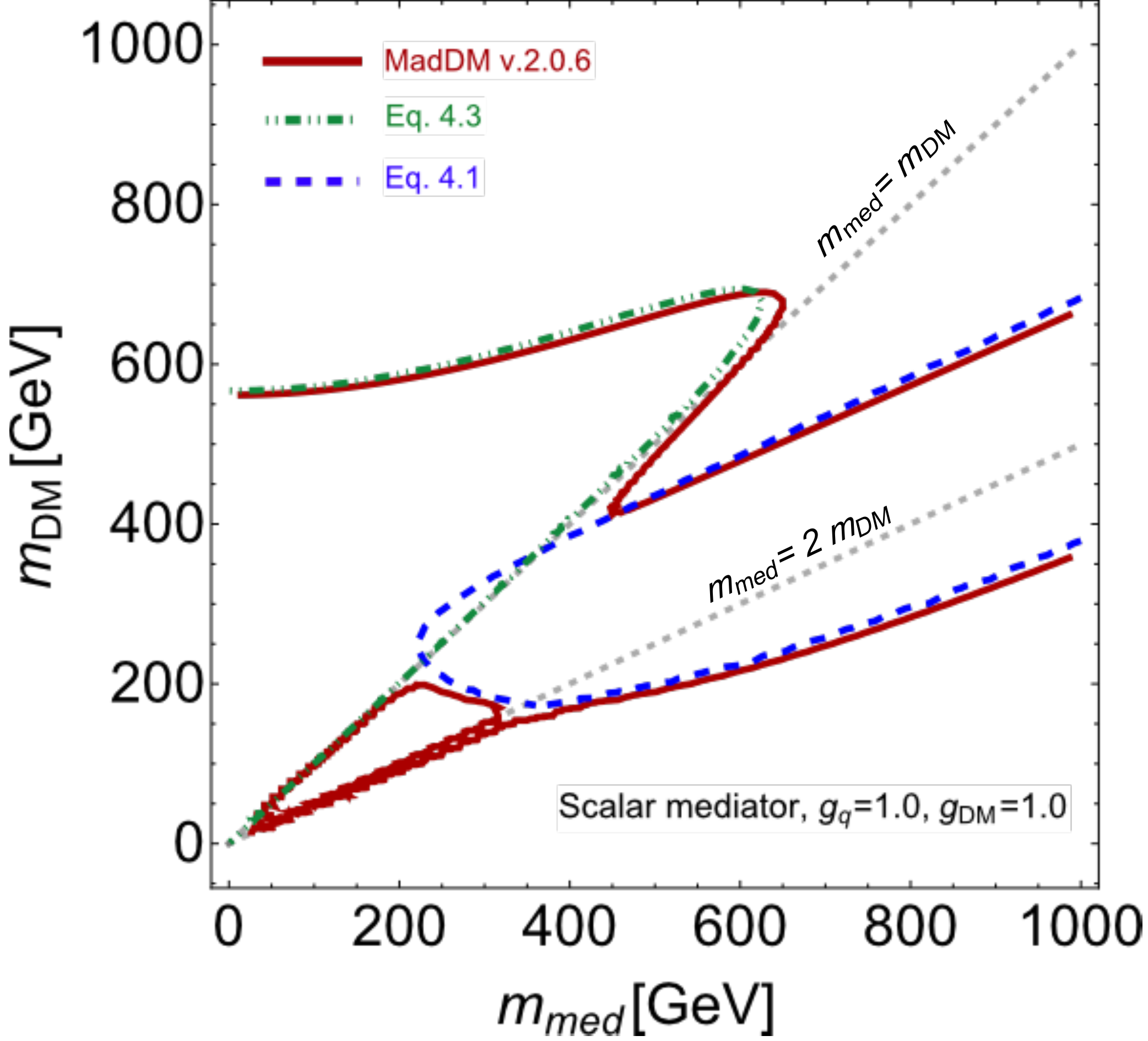}
\includegraphics[width=0.47\textwidth]{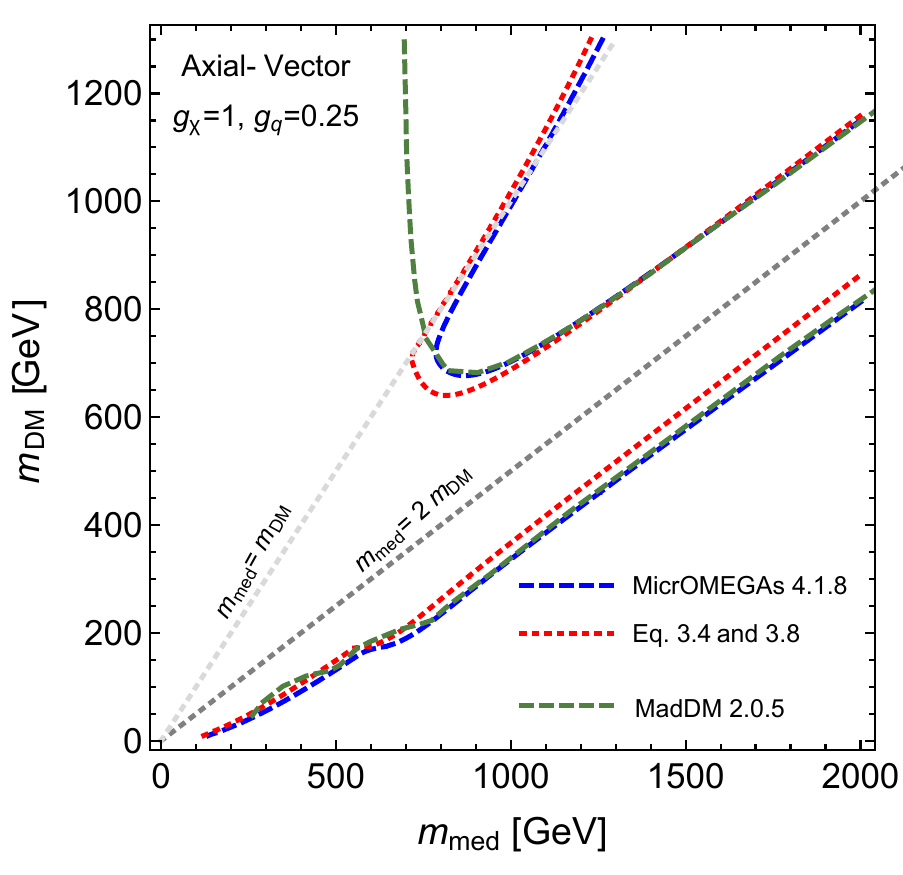} 
\caption{Comparison of the analytic calculations described in Section~\ref{analyticrelic}  with numerical results obtained with \maddm and \textsc{MicrOMEGAs}. In the scalar case (left panel) the exact numerical results have been obtained with version 2.0.6 of \maddm, while for the axial-vector mediator model (right panel)
version 4.1.8 of \textsc{MicrOMEGAs} has been used to calculate the relic density. See main text for further details.}
\label{fig:analcalc}
\end{figure}
\end{center}

In Figure~\ref{fig:analcalc} we compare the results of the analytic calculations described in Section~\ref{analyticrelic}  with full numerical results obtained with the packages \maddm and \textsc{MicrOMEGAs}~\cite{Belanger:2014vza}. The left~(right) panel shows the contours in the $M_{\rm med}$--$m_{\rm DM}$ plane for which $\Omega h^2 = 0.118$ in the case of scalar (axial-vector) interactions for the coupling choices $g_{\rm DM} = 1$ and $g_q = 1$~($g_q = 0.25$). The analytic results are obtained by employing~(\ref{eq31}) and (\ref{eq35}) in the case of the scalar mediator, while (\ref{eq34}) and (\ref{eq38}) are used for the axial-vector simplified model. In both cases the ratio of the DM mass to freeze-out temperature is fixed to~$x_f = 28$ and also the effective number of relativistic degrees of freedom is kept constant across the~$M_{\rm med}$--$m_{\rm DM}$ plane.

From both panels it is evident that in the limiting cases where one of the channels dominates the annihilation cross sections, the analytic calculations are in good agreement with the numerical results obtained either by version 2.0.6 of \maddm (scalar case) or version 4.1.8 of \textsc{MicrOMEGAs} (axial-vector case). We emphasize that an improved agreement between analytic and numerical calculations can be obtained when using thermally-averaged cross sections and a numerically determined freeze-out temperature~\cite{Gondolo:1990dk}. In the case of the axial-vector model, we also display the contour line with $\Omega h^2 = 0.118$ using version 2.0.5 of \maddm. This prediction is used as representative of the results of \cite{Pree:2016hwc} that did not incorporate $t$-channel annihilation contributions. One observes that not taking into account the annihilation  contribution (\ref{eq38}) leads to an erroneous prediction of overabundance for $M_{\rm med} < m_{\rm DM}$. We have also verified that version 2.0.6 of \maddm and version 4.1.8 of \textsc{MicrOMEGAs} lead to compatible result for axial-vector simplified models. 

%%%%%%%%%%%%%%%%

\section{Conclusions}

Simplified DM models in which the interactions between the visible and the invisible sector are mediated by the exchange of TeV-scale particles represent an interesting class of all  possible realizations of physics beyond the SM with a  viable DM candidate. In such scenarios the decays of the mediators can lead to final states where DM is  accompanied by SM radiation (so-called mono-$X$ events) but also to signatures that feature only SM particles (such as for instance di-jet or di-lepton events). To map out the DM parameter space in a given simplified model comparing and combining the different LHC search strategies is an important task. 

This document describes simplified spin-1 models where a $s$-channel mediator couple to DM, quarks, and leptons. Four benchmark scenarios with  different relative sizes of quark and lepton couplings are recommended that exemplify the rich phenomenology of the simplified DM model in the mono-jet, di-jet, and di-lepton channels. In this document, benchmark points for the vector and axial-vector mediator models where the mediator width is large have not been considered. Even though these may evade the constraints provided by di-jet and di-lepton resonance searches, LHC searches for non-resonant phenomena in the same final states, such as the ones in~\cite{ATLAS:2015nsi,CMS-PAS-EXO-15-009,Aad:2014wca, Khachatryan:2015scf}, are sensitive to wide mediators and can be subject of dedicated future studies.

This document also presents improved numerical calculations for the relic density for the dominant annihilation processes  that  involve  spin-0  and  spin-1  mediators. The full numerical results are compared to analytical calculations and found to be in good agreement in all cases  where the annihilation cross section is dominated by a single channel. It is furthermore shown that the omission of $t$-channel annihilation contributions can lead to parameter regions which feature an  erroneous  relic overabundance or  underabundance. 

\section{Acknowledgement} 
The work of CD is supported by the European Research Council under the European Union's Horizon 2020 research and innovation programme (ERC grant agreement No. 679305 DARKJETS).  MF and PT are funded by the European Research Council under the European Union's Horizon 2020 program (ERC grant agreement No. 648680 DARKHORIZONS). UH acknowledges the hospitality and support of the CERN theory division. The work of GL is partially supported by the DOE Award DE-SC0010010. AR is supported by the Swiss National Science Foundation (SNSF), project title: ``Investigating the Nature of Dark Matter'', project number: 200020-159223. The work of TMPT is supported in part by National Science Foundation grants PHY-1316792 and PHY-1620638. 

\bibliography{ref}

\providecommand{\href}[2]{#2}\begingroup\raggedright\begin{thebibliography}{10}

\bibitem{Abercrombie:2015wmb}
D.~Abercrombie et~al., {\it {Dark Matter Benchmark Models for Early LHC Run-2
  Searches: Report of the ATLAS/CMS Dark Matter Forum}},
  \href{http://arxiv.org/abs/1507.00966}{{\tt arXiv:1507.00966}}.

\bibitem{Dreiner:2013vla}
H.~Dreiner, D.~Schmeier, and J.~Tattersall, {\it {Contact Interactions Probe
  Effective Dark Matter Models at the LHC}},  {\em Europhys. Lett.} {\bf 102}
  (2013) 51001, [\href{http://arxiv.org/abs/1303.3348}{{\tt arXiv:1303.3348}}].

\bibitem{Chala:2015ama}
M.~Chala, F.~Kahlhoefer, M.~McCullough, G.~Nardini, and K.~Schmidt-Hoberg, {\it
  {Constraining Dark Sectors with Monojets and Dijets}},  {\em JHEP} {\bf 07}
  (2015) 089, [\href{http://arxiv.org/abs/1503.05916}{{\tt arXiv:1503.05916}}].

\bibitem{Fairbairn:2016iuf}
M.~Fairbairn, J.~Heal, F.~Kahlhoefer, and P.~Tunney, {\it {Constraints on Z′
  models from LHC dijet searches and implications for dark matter}},  {\em
  JHEP} {\bf 09} (2016) 018, [\href{http://arxiv.org/abs/1605.07940}{{\tt
  arXiv:1605.07940}}].

\bibitem{ATLAS:2016pyq}
A.~Collaboration, {\it {Search for heavy Higgs bosons A/H decaying to a
  top-quark pair in pp collisions at $\sqrt{s}=8$ TeV with the ATLAS
  detector}}, .

\bibitem{Bauer:2017ota}
M.~Bauer, U.~Haisch, and F.~Kahlhoefer, {\it {Simplified dark matter models
  with two Higgs doublets: I. Pseudoscalar mediators}},
  \href{http://arxiv.org/abs/1701.07427}{{\tt arXiv:1701.07427}}.

\bibitem{Boveia:2016mrp}
G.~Busoni et~al., {\it {Recommendations on presenting LHC searches for missing
  transverse energy signals using simplified $s$-channel models of dark
  matter}},  \href{http://arxiv.org/abs/1603.04156}{{\tt arXiv:1603.04156}}.

\bibitem{ATLASsummaryplots}
{ATLAS Collaboration}, ``Dark matter simplified model exclusions.''
  \href{https://atlas.web.cern.ch/Atlas/GROUPS/PHYSICS/CombinedSummaryPlots/EXOTICS/ATLAS\_DarkMatter\_Summary/history.html}{link},
  2016.

\bibitem{CMS_SummaryPlots_ICHEP}
{CMS Collaboration}, ``{Dark Matter Summary Plots from CMS for ICHEP 2016}.''
  \href{https://cds.cern.ch/record/2208044/files/DP2016\_057.pdf}{CMS-DP-16-057}.

\bibitem{Sirunyan:2016iap}
{CMS Collaboration}, {\it {Search for dijet resonances in proton-proton
  collisions at $\sqrt{s}=$ 13 TeV and constraints on dark matter and other
  models}},  {\em Submitted to: Phys. Lett. B} (2016)
  [\href{http://arxiv.org/abs/1611.03568}{{\tt arXiv:1611.03568}}].

\bibitem{Kahlhoefer:2015bea}
F.~Kahlhoefer, K.~Schmidt-Hoberg, T.~Schwetz, and S.~Vogl, {\it {Implications
  of unitarity and gauge invariance for simplified dark matter models}},  {\em
  JHEP} {\bf 02} (2016) 016, [\href{http://arxiv.org/abs/1510.02110}{{\tt
  arXiv:1510.02110}}].

\bibitem{Jacques:2016dqz}
T.~Jacques, A.~Katz, E.~Morgante, D.~Racco, M.~Rameez, and A.~Riotto, {\it
  {Complementarity of DM searches in a consistent simplified model: the case of
  $Z′$}},  {\em JHEP} {\bf 10} (2016) 071,
  [\href{http://arxiv.org/abs/1605.06513}{{\tt arXiv:1605.06513}}].

\bibitem{Arcadi:2013qia}
G.~Arcadi, Y.~Mambrini, M.~H.~G. Tytgat, and B.~Zaldivar, {\it {Invisible
  $Z^\prime$ and dark matter: LHC vs LUX constraints}},  {\em JHEP} {\bf 1403}
  (2014) 134, [\href{http://arxiv.org/abs/1401.0221}{{\tt arXiv:1401.0221}}].

\bibitem{Bauer:2016gys}
M.~Bauer et~al., {\it {Towards the next generation of simplified Dark Matter
  models}},  \href{http://arxiv.org/abs/1607.06680}{{\tt arXiv:1607.06680}}.

\bibitem{Carena:2004xs}
M.~Carena, A.~Daleo, B.~A. Dobrescu, and T.~M.~P. Tait, {\it {$Z^\prime$ gauge
  bosons at the Tevatron}},  {\em Phys. Rev.} {\bf D70} (2004) 093009,
  [\href{http://arxiv.org/abs/hep-ph/0408098}{{\tt hep-ph/0408098}}].

\bibitem{D'Ambrosio:2002ex}
G.~D'Ambrosio, G.~F. Giudice, G.~Isidori, and A.~Strumia, {\it {Minimal flavor
  violation: An Effective field theory approach}},  {\em Nucl. Phys.} {\bf
  B645} (2002) 155--187, [\href{http://arxiv.org/abs/hep-ph/0207036}{{\tt
  hep-ph/0207036}}].

\bibitem{LHCDMFmodels}
``{Model Repository of the ATLAS/CMS Dark Matter Forum}.''
  \href{svn+ssh://svn.cern.ch/reps/LHCDMF}{svn+ssh://svn.cern.ch/reps/LHCDMF},
  2015.

\bibitem{Alwall:2014hca}
J.~Alwall, R.~Frederix, S.~Frixione, V.~Hirschi, F.~Maltoni, O.~Mattelaer,
  H.~S. Shao, T.~Stelzer, P.~Torrielli, and M.~Zaro, {\it {The automated
  computation of tree-level and next-to-leading order differential cross
  sections, and their matching to parton shower simulations}},  {\em JHEP} {\bf
  07} (2014) 079, [\href{http://arxiv.org/abs/1405.0301}{{\tt
  arXiv:1405.0301}}].

\bibitem{Backovic:2015soa}
M.~Backovic, M.~Kr{\"a}mer, F.~Maltoni, A.~Martini, K.~Mawatari, and M.~Pellen,
  {\it {Higher-order QCD predictions for dark matter production at the LHC in
  simplified models with s-channel mediators}},  {\em Eur. Phys. J.} {\bf C75}
  (2015), no.~10 482, [\href{http://arxiv.org/abs/1508.05327}{{\tt
  arXiv:1508.05327}}].

\bibitem{Degrande:2011ua}
C.~Degrande, C.~Duhr, B.~Fuks, D.~Grellscheid, O.~Mattelaer, and T.~Reiter,
  {\it {UFO - The Universal FeynRules Output}},  {\em Comput. Phys. Commun.}
  {\bf 183} (2012) 1201--1214, [\href{http://arxiv.org/abs/1108.2040}{{\tt
  arXiv:1108.2040}}].

\bibitem{Alloul:2013bka}
A.~Alloul, N.~D. Christensen, C.~Degrande, C.~Duhr, and B.~Fuks, {\it
  {FeynRules 2.0 - A complete toolbox for tree-level phenomenology}},  {\em
  Comput. Phys. Commun.} {\bf 185} (2014) 2250--2300,
  [\href{http://arxiv.org/abs/1310.1921}{{\tt arXiv:1310.1921}}].

\bibitem{DMsimp}
``{Simplified dark matter models}.''
  \href{http://feynrules.irmp.ucl.ac.be/wiki/DMsimp}{http://feynrules.irmp.ucl.ac.be/wiki/DMsimp}.

\bibitem{Duerr:2016tmh}
M.~Duerr, F.~Kahlhoefer, K.~Schmidt-Hoberg, T.~Schwetz, and S.~Vogl, {\it {How
  to save the WIMP: global analysis of a dark matter model with two s-channel
  mediators}},  {\em JHEP} {\bf 09} (2016) 042,
  [\href{http://arxiv.org/abs/1606.07609}{{\tt arXiv:1606.07609}}].

\bibitem{Aad:2014cka}
{ATLAS~Collaboration}, {\it {Search for high-mass dilepton resonances in pp
  collisions at $\sqrt{s}=8$  TeV with the ATLAS detector}},  {\em Phys.
  Rev.} {\bf D90} (2014), no.~5 052005,
  [\href{http://arxiv.org/abs/1405.4123}{{\tt arXiv:1405.4123}}].

\bibitem{Khachatryan:2014fba}
C.~Collaboration, {\it {Search for physics beyond the standard model in
  dilepton mass spectra in proton-proton collisions at $ \sqrt{s}=8 $ TeV}},
  {\em JHEP} {\bf 04} (2015) 025, [\href{http://arxiv.org/abs/1412.6302}{{\tt
  arXiv:1412.6302}}].

\bibitem{Aaboud:2016cth}
{ATLAS~Collaboration}, {\it {Search for high-mass new phenomena in the dilepton
  final state using proton-proton collisions at $\sqrt{s}=13$ TeV with the
  ATLAS detector}},  {\em Phys. Lett.} {\bf B761} (2016) 372--392,
  [\href{http://arxiv.org/abs/1607.03669}{{\tt arXiv:1607.03669}}].

\bibitem{Khachatryan:2016zqb}
{CMS~Collaboration}, {\it {Search for narrow resonances in dilepton mass
  spectra in proton-proton collisions at $\sqrt{s}$ = 13 TeV and combination
  with 8 TeV data}},  \href{http://arxiv.org/abs/1609.05391}{{\tt
  arXiv:1609.05391}}.

\bibitem{Khachatryan:2016qkc}
{CMS Collaboration}, {\it {Search for heavy resonances decaying to tau lepton
  pairs in proton-proton collisions at $\sqrt{s}$ = 13 TeV}},  {\em Submitted
  to: JHEP} (2016) [\href{http://arxiv.org/abs/1611.06594}{{\tt
  arXiv:1611.06594}}].

\bibitem{ATLAS-CONF-2016-045}
{ATLAS Collaboration}, {\it {Search for new high-mass resonances in the
  dilepton final state using proton-proton collisions at $\sqrt{s}$ = 13 TeV
  with the ATLAS detector}},  Tech. Rep. ATLAS-CONF-2016-045, CERN, Geneva,
  Aug, 2016.

\bibitem{Dicus:1994bm}
D.~Dicus, A.~Stange, and S.~Willenbrock, {\it {Higgs decay to top quarks at
  hadron colliders}},  {\em Phys. Lett.} {\bf B333} (1994) 126--131,
  [\href{http://arxiv.org/abs/hep-ph/9404359}{{\tt hep-ph/9404359}}].

\bibitem{Frederix:2007gi}
R.~Frederix and F.~Maltoni, {\it {Top pair invariant mass distribution: A
  Window on new physics}},  {\em JHEP} {\bf 01} (2009) 047,
  [\href{http://arxiv.org/abs/0712.2355}{{\tt arXiv:0712.2355}}].

\bibitem{Djouadi:2015jea}
A.~Djouadi, L.~Maiani, A.~Polosa, J.~Quevillon, and V.~Riquer, {\it {Fully
  covering the MSSM Higgs sector at the LHC}},  {\em JHEP} {\bf 06} (2015) 168,
  [\href{http://arxiv.org/abs/1502.05653}{{\tt arXiv:1502.05653}}].

\bibitem{Craig:2015jba}
N.~Craig, F.~D'Eramo, P.~Draper, S.~Thomas, and H.~Zhang, {\it {The Hunt for
  the Rest of the Higgs Bosons}},  {\em JHEP} {\bf 06} (2015) 137,
  [\href{http://arxiv.org/abs/1504.04630}{{\tt arXiv:1504.04630}}].

\bibitem{Jung:2015gta}
S.~Jung, J.~Song, and Y.~W. Yoon, {\it {Dip or nothingness of a Higgs resonance
  from the interference with a complex phase}},  {\em Phys. Rev.} {\bf D92}
  (2015), no.~5 055009, [\href{http://arxiv.org/abs/1505.00291}{{\tt
  arXiv:1505.00291}}].

\bibitem{Bernreuther:2015fts}
W.~Bernreuther, P.~Galler, C.~Mellein, Z.~G. Si, and P.~Uwer, {\it {Production
  of heavy Higgs bosons and decay into top quarks at the LHC}},  {\em Phys.
  Rev.} {\bf D93} (2016), no.~3 034032,
  [\href{http://arxiv.org/abs/1511.05584}{{\tt arXiv:1511.05584}}].

\bibitem{Gori:2016zto}
S.~Gori, I.-W. Kim, N.~R. Shah, and K.~M. Zurek, {\it {Closing the Wedge:
  Search Strategies for Extended Higgs Sectors with Heavy Flavor Final
  States}},  {\em Phys. Rev.} {\bf D93} (2016), no.~7 075038,
  [\href{http://arxiv.org/abs/1602.02782}{{\tt arXiv:1602.02782}}].

\bibitem{Carena:2016npr}
M.~Carena and Z.~Liu, {\it {Challenges and opportunities for heavy scalar
  searches in the $ t\overline{t} $ channel at the LHC}},  {\em JHEP} {\bf 11}
  (2016) 159, [\href{http://arxiv.org/abs/1608.07282}{{\tt arXiv:1608.07282}}].

\bibitem{Ade:2015xua}
{\bf Planck} Collaboration, P.~A.~R. Ade et~al., {\it {Planck 2015 results.
  XIII. Cosmological parameters}},  \href{http://arxiv.org/abs/1502.01589}{{\tt
  arXiv:1502.01589}}.

\bibitem{Backovic:2013dpa}
M.~Backovic, K.~Kong, and M.~McCaskey, {\it {MadDM v.1.0: Computation of Dark
  Matter Relic Abundance Using MadGraph5}},  {\em Physics of the Dark Universe}
  {\bf 5-6} (2014) 18--28, [\href{http://arxiv.org/abs/1308.4955}{{\tt
  arXiv:1308.4955}}].

\bibitem{Backovic:2015tpt}
M.~Backovic, A.~Martini, K.~Kong, O.~Mattelaer, and G.~Mohlabeng, {\it {MadDM:
  New dark matter tool in the LHC era}},  {\em AIP Conf. Proc.} {\bf 1743}
  (2016) 060001, [\href{http://arxiv.org/abs/1509.03683}{{\tt
  arXiv:1509.03683}}].

\bibitem{RelicRepo}
D.~M.~W. Group, ``Relic density curves.''
  \url{https://gitlab.cern.ch/lhc-dmwg-material/relic-density}, 2017.

\bibitem{Abdullah:2014lla}
M.~Abdullah, A.~DiFranzo, A.~Rajaraman, T.~M.~P. Tait, P.~Tanedo, and A.~M.
  Wijangco, {\it {Hidden on-shell mediators for the Galactic Center
  $\gamma$-ray excess}},  {\em Phys. Rev.} {\bf D90} (2014) 035004,
  [\href{http://arxiv.org/abs/1404.6528}{{\tt arXiv:1404.6528}}].

\bibitem{Busoni:2014gta}
G.~Busoni, A.~De~Simone, T.~Jacques, E.~Morgante, and A.~Riotto, {\it {Making
  the Most of the Relic Density for Dark Matter Searches at the LHC 14 TeV
  Run}},  {\em JCAP} {\bf 1503} (2015), no.~03 022,
  [\href{http://arxiv.org/abs/1410.7409}{{\tt arXiv:1410.7409}}].

\bibitem{Pree:2016hwc}
T.~du~Pree, K.~Hahn, P.~Harris, and C.~Roskas, {\it {Cosmological constraints
  on Dark Matter models for collider searches}},
  \href{http://arxiv.org/abs/1603.08525}{{\tt arXiv:1603.08525}}.

\bibitem{Belanger:2014vza}
G.~Belanger, F.~Boudjema, A.~Pukhov, and A.~Semenov, {\it {micrOMEGAs4.1: two
  dark matter candidates}},  {\em Comput. Phys. Commun.} {\bf 192} (2015)
  322--329, [\href{http://arxiv.org/abs/1407.6129}{{\tt arXiv:1407.6129}}].

\bibitem{Gondolo:1990dk}
P.~Gondolo and G.~Gelmini, {\it {Cosmic abundances of stable particles:
  Improved analysis}},  {\em Nucl.Phys.} {\bf B360} (1991) 145--179.

\bibitem{ATLAS:2015nsi}
{ATLAS Collaboration}, {\it {Search for new phenomena in dijet mass and angular
  distributions from $pp$ collisions at $\sqrt{s}=$ 13 TeV with the ATLAS
  detector}},  {\em Phys. Lett.} {\bf B754} (2016) 302--322,
  [\href{http://arxiv.org/abs/1512.01530}{{\tt arXiv:1512.01530}}].

\bibitem{CMS-PAS-EXO-15-009}
{CMS Collaboration}, {\it {Searches for quark contact interactions and extra
  spatial dimensions with dijet angular distributions in proton proton
  collisions at 13 TeV}},  Tech. Rep. CMS-PAS-EXO-15-009, CERN, Geneva, 2015.

\bibitem{Aad:2014wca}
{ATLAS Collaboration}, {\it {Search for contact interactions and large extra
  dimensions in the dilepton channel using proton-proton collisions at
  $\sqrt{s}$ = 8 TeV with the ATLAS detector}},  {\em Eur. Phys. J.} {\bf C74}
  (2014), no.~12 3134, [\href{http://arxiv.org/abs/1407.2410}{{\tt
  arXiv:1407.2410}}].

\bibitem{Khachatryan:2015scf}
V.~Khachatryan et~al., {\it {Search for excited leptons in proton-proton
  collisions at sqrt(s) = 8 TeV}},  {\em JHEP} {\bf 03} (2016) 125,
  [\href{http://arxiv.org/abs/1511.01407}{{\tt arXiv:1511.01407}}].

\end{thebibliography}\endgroup
\bibliographystyle{JHEP}

\end{document}